\documentclass[journal]{IEEEtran}

\usepackage{enumerate}
\usepackage{graphicx}
\usepackage{epsf}
\usepackage{subfigure}
\usepackage{amsmath}
\usepackage{amssymb}
\usepackage{array}
\usepackage{setspace}
\usepackage[amsmath,thmmarks]{ntheorem}
\usepackage[ruled,lined]{algorithm2e}
\usepackage{algorithmic}
\usepackage{color}
\usepackage{amsfonts}
\usepackage{dsfont}
\usepackage{xfrac}
\usepackage{breqn}
\usepackage{bbm}
\usepackage{cite}

\graphicspath{{Fig/},{fig/}}

\theoremseparator{.}

\begin{document}

\title{\huge Energy Efficient Active Stacked Intelligent Metasurfaces}


\author{
Li-Hsiang Shen,~\IEEEmembership{Member,~IEEE}}

\maketitle

\begin{abstract}
This paper investigates an energy-efficient active stacked intelligent metasurfaces (ASIM)-assisted downlink transmission framework, where a multi-antenna base station (BS) serves multiple users through a multi-layer metasurface architecture. Unlike conventional passive intelligent surfaces, the considered ASIM employs active amplification and multiple transmissive layers to enhance electromagnetic wave manipulation. We aim to maximize the system energy efficiency (EE) by jointly optimizing the BS beamforming and ASIM configurations under user quality-of-service and amplification constraints. The resulting problem is highly coupled and non-convex due to the cascaded near-field channel and multi-layer metasurface structure. To address this challenge, we first transform the original problem through Dinkelbach, Lagrangian dual, and quadratic transformations. An alternative optimization framework is then developed, where the BS beamforming subproblem is solved via successive convex approximation (SCA), while the ASIM configuration is optimized using Bayesian optimization based on a Gaussian process surrogate model. Numerical results demonstrate that the proposed scheme significantly improves the achievable EE compared to conventional passive SIM and heuristic benchmark methods. Furthermore, the impacts of amplification capability, number of metasurface layers, and inter-layer spacing on system performance are investigated, providing useful design insights for future active metasurface-assisted wireless networks.
\end{abstract}
\begin{IEEEkeywords}
Active stacked intelligent metasurfaces, energy efficiency, successive convex approximation, Bayesian optimization.
\end{IEEEkeywords}

%

{\let\thefootnote\relax\footnotetext
{Li-Hsiang Shen is with the Department of Communication Engineering, National Central University, Taoyuan 320317, Taiwan. (email: shen@ncu.edu.tw)}}

\section{Introduction}

Reconfigurable intelligent surfaces (RISs) have recently emerged as a promising technology for shaping wireless propagation environments through programmable electromagnetic responses \cite{shen_acm, wu2019towards}. By properly adjusting the phase-shifts of a large number of passive meta-elements, RISs can enhance signal quality, extend network coverage, and improve spectral efficiency without requiring additional radio-frequency chains \cite{shen_amy}. However, conventional RIS architectures suffer from the multiplicative fading channel effect, where the cascaded transmitter-RIS-user channel experiences severe pathloss \cite{basar2019wireless, my_star}, particularly in long-distance transmissions and high-frequency bands.

To overcome this limitation, active intelligent surfaces have been proposed by integrating amplification circuits into meta-elements \cite{my_ps, a_ris, my_mfris}, allowing the reflected or transmitted signals to be amplified while performing electromagnetic manipulation. More recently, research on stacked intelligent metasurfaces (SIM) \cite{sim1, sim2, sim3} has attracted significant attention thanks to their capability of establishing multiple electromagnetic processing layers. By introducing multiple closely spaced transmissive RIS layers, SIM can exploit near-field wave coupling among layers to realize enhanced beam focusing and wavefront shaping beyond conventional single-layer RISs \cite{sim4}. Unlike conventional active RIS consisting of a single-layer active metasurface, the proposed active SIM (ASIM) employs multiple cascaded active transmissive metasurface layers. The resulting architecture exploits near-field electromagnetic coupling among adjacent layers while simultaneously optimizing amplification gains and phase-shifts across all layers. Compared to passive SIM \cite{sim1, sim2, sim3, eesim}, ASIM further introduces active amplification, which alleviates the severe propagation loss of cascaded channels at the expense of additional circuit power consumption and amplified noise. Additionally, \cite{eesim} points out that determining the optimal number of SIM layers is essential for maximizing energy efficiency (EE). However, the study only considers passive SIM architectures. From an implementation perspective, each ASIM layer consists of programmable active meta-atoms, where the power gain and phase shift of each element are independently controlled through dedicated bias voltages generated by a microcontroller. Such an architecture is consistent with existing active RIS implementations.

Despite their potential, optimizing ASIM remains highly challenging. The cascaded near-field channel among stacked layers introduces strong coupling among amplification coefficients and phase-shifts across different layers, resulting in a highly nonlinear and non-convex optimization problem. Moreover, increasing the amplification gain can improve the achievable rate but simultaneously enlarges power consumption and amplified noise propagation. Therefore, the tradeoff between communication performance and energy expenditure becomes a critical issue. Furthermore, conventional alternating optimization and layer-by-layer optimization \cite{sim_iter} approaches often suffer from excessive computational complexity and local optima due to the high dimensionality of ASIM configurations. Motivated by these challenges, we investigate an energy-efficient ASIM-assisted downlink system by jointly optimizing the BS beamforming and multi-layer ASIM configurations. The main contributions of this work are summarized as follows:
\begin{itemize}
	\item We have conceived the energy-efficient optimization framework of ASIM, jointly modeling near-field inter-layer coupling, active amplification, amplification noise accumulation, and practical circuit power consumption. This model extends existing passive SIM and single-layer active RIS models to a multi-layer active structure. We formulate a joint BS beamforming and ASIM configuration optimization problem for maximizing the system EE under transmit power, amplification, and quality-of-service (QoS) constraints.
	
	\item We develop an alternative optimization (AO) framework that decomposes the original problem into BS beamforming and ASIM configuration subproblems. The beamforming design is solved via Dinkelbach transformation, quadratic transformation, and successive convex approximation (SCA). A Bayesian optimization-based ASIM configuration algorithm is designed, which exploits a Gaussian process surrogate model to efficiently optimize the joint amplification and phase-shifts across all ASIM layers.
	
	\item Numerical results demonstrate the superiority of the proposed design over conventional passive SIM schemes and reveal the impacts of amplification capability, number of layers, inter-layer spacing, and power consumption on system EE.
\end{itemize}

\begin{figure}[!t]
\centering
\includegraphics[width=3in]{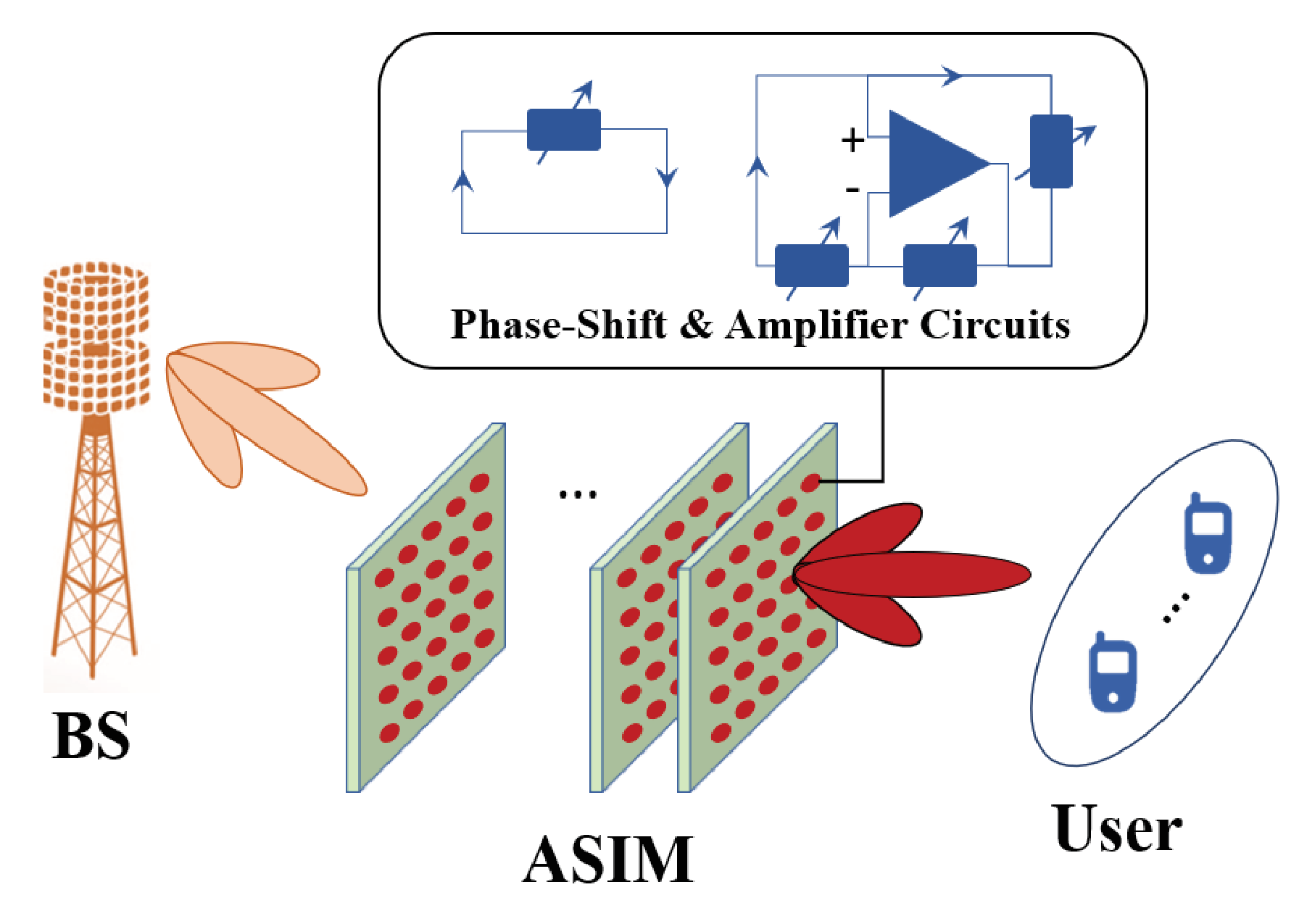}
\caption{The proposed ASIM architecture in multiuser downlink transmissions.} \label{architecture}
\end{figure}

\section{System Model and Problem Formulation}

In Fig. \ref{architecture}, we consider a BS equipped with $N$ transmitter (Tx) antennas, with its set of $\mathcal{N} = \{1, ...,n,..., N\}$, serving $K$ downlink (DL) users equipped with a single antenna, with its set $\mathcal{K} = \{1, ...,k,..., K\}$. Note that the users are located beyond the ASIM. We consider that one $L$-layer ASIM is deployed to alternate the channel, with the set of $\mathcal{L} = \{1,...,l,...,L\}$. Each layer has the $M$ elements, associated with the set of $\mathcal{M} = \{1,...,m,...,M\}$. We define the channel from Tx to layer 1 of ASIM as $\mathbf{H}_{0}$ and from layer $L$ to DL user as $\mathbf{g}_{k}$. Note that no direct link is considered due to potential blockages. We consider the Rician channel fading between the BS and the layer 1 of ASIM, which is given by
\begin{align} \label{channel}
	\mathbf{H}_{0} = \sqrt{h_0 d_{0}^{-\alpha}} \left( \sqrt{\frac{\kappa}{1+\kappa}}\mathbf{H}^{\text{LoS}}_{0} + \sqrt{\frac{1}{1+\kappa}}\mathbf{H}^{\text{NLoS}}_{0} \right),
\end{align}
where $h_0$ is the reference signal strength at 1 meter, $d_{0}$ is the distance between the BS and layer 1 of ASIM, and $\alpha$ indicates the pathloss exponent. Notation $\kappa$ means the Rician factors. Consider the planar array for both BS and ASIM with their size $N = N_{h} N_{v}$ and $M = M_h M_v$, where $h$ and $v$ indicate the horizontal/vertical axis of the array, respectively. The LoS components is given by $\mathbf{H}^{\text{LoS}}_{0} = \mathbf{a}_{\text{S}}(\theta^{\text{AoA}}_{\text{S}},\phi^{\text{AoA}}_{\text{S}}) \mathbf{a}_{\text{B}}^{\mathcal{H}}(\theta^{\text{AoD}}_{\text{B}}, \phi^{\text{AoD}}_{\text{B}}) \in \mathbb{C}^{M\times N}$, where the component of the steering vector of ASIM and of BS are respectively given by
\begingroup
\allowdisplaybreaks
\begin{align} \label{array_ris}
	& [\mathbf{a}_{\text{S}} (\theta^{\text{AoA}}_{\text{S}},\phi^{\text{AoA}}_{\text{S}})]_{m_h, m_v} = \frac{1}{\sqrt{M}} \cdot \notag\\
	&  \qquad\qquad e^{j\frac{2\pi d_{\text{S}}}{\lambda} 
	\left( 
	m_h \sin \theta_{\text{S}}^{\text{AoA}} \cos \phi_{\text{S}}^{\text{AoA}} +
	m_v \sin \theta_{\text{S}}^{\text{AoA}} \sin \phi_{\text{S}}^{\text{AoA}}
	\right)
	}, \\
	& [\mathbf{a}_{\text{B}}(\theta^{\text{AoD}}_{\text{B}},\phi^{\text{AoD}}_{\text{B}})]_{n_h,n_v} = \frac{1}{\sqrt{N}} \cdot  \notag\\
	& \qquad\qquad e^{j\frac{2\pi d_{\text{B}}}{\lambda} 
	\left( 
	n_{h} \sin \theta_{\text{B}}^{\text{AoD}} \cos \phi_{\text{B}}^{\text{AoD}} +
	n_{v} \sin \theta_{\text{B}}^{\text{AoD}} \sin \phi_{\text{B}}^{\text{AoD}}
	\right)
	}, \label{array_bs}
\end{align} 
\endgroup
where their angle-of-departures (AoD) and angle-of-arrivals (AoA) in azimuth and elevation are denoted as $\{\theta^{\text{AoA}}_{\text{S}},\phi^{\text{AoA}}_{\text{S}}, \theta^{\text{AoD}}_{\text{B}},\phi^{\text{AoD}}_{\text{B}} \}$. $\mathcal{H}$ is Hermitian operation. While, $\mathbf{H}^{\text{NLoS}}$ indicates the NLoS channel following Rayleigh fading. Similar to \eqref{channel}, we consider the Rician channel between the layer $L$ of ASIM and the DL user $k$, represented by
\begin{align} \label{channel_ue}
	\mathbf{g}_{k} \!=\! \sqrt{ h_{0,k} d_{k}^{-\alpha_x}} \left( \sqrt{\frac{\kappa_k}{1 + \kappa_k}} \mathbf{g}^{\text{LoS}}_{k} + \sqrt{\frac{1}{1+\kappa_k}}\mathbf{g}^{\text{NLoS}}_{k} \right),
\end{align}
The definitions of pertinent parameters of \eqref{channel_ue} are the same as those in \eqref{channel}. Note that $\mathbf{g}^{\text{LoS}}_{k} \in \mathbb{C}^{M \times 1}$ follows \eqref{array_ris} associated with layer $L$ of ASIM, which is neglected here for simplicity.

For the ASIM channel, it possesses only transmission capability for DL users. The effective channel from the BS Tx, ASIM, to the DL user $k$ can be given by
\begin{align} \label{ch_general}
	\mathbf{h}_{k} = \mathbf{g}_{k}^{\mathcal{H}} \boldsymbol{\Theta}_{L} \mathbf{H}_{L-1} \boldsymbol{\Theta}_{L-1} ...\boldsymbol{\Theta}_{2} \mathbf{H}_{1} \boldsymbol{\Theta}_{1} \mathbf{H}_{0},
\end{align}
whereas the intermediate channel from layer $l$ to layer $l+1$ is denoted as $\mathbf{H}_{l}$. Notation $\boldsymbol{\Theta}_{l} = \operatorname{diag}( \varphi_{l,m}) = \operatorname{diag}( \sqrt{\beta_{l,m}} e^{j\theta_{l,m}})$ indicates the layer $l$'s configuration of ASIM, where $\beta_{l,m}\in[0, \beta_{\max}]$ and $\theta_{l,m}\in [0, 2\pi)$ denote the active power gain and phase-shifts of element $m$ at ASIM layer $l$, respectively. We further notice that $\beta_{\max}\leq 1$ indicates the conventional passive SIM without amplification capability, whereas $\beta_{\max}>1$ stands for the configuration of amplified SIM structure. According to the Rayleigh-Sommerfeld diffraction theory \cite{sim2}, the channel coefficient from the element $m$ at layer $l$ to element $m'$ at next layer $l+1$ is expressed as
\begin{align}
	h_{m,m'}^{l} = \frac{A_m \cos \chi_{m,m'}^{l}}{r_{m,m'}^{l}} \left( \frac{1}{2\pi r_{m,m'}^{l}} - j\frac{1}{\lambda} \right) e^{j 2 \pi \frac{ r_{m,m'}^{l} }{\lambda}},
\end{align}
where $r_{m,m'}^{l}$ indicates the corresponding transmission distance between the inter-layer elements, $A_m$ is the area of each element of ASIM, and $\chi_{m,m'}^{l}$ represents the angle between the propagation direction and the normal direction of the transmissive layer. We further notice that the inter-layer distance of ASIM equally deployed is defined as $d$.

The received signal of user $k$ is expressed as $s_k = \mathbf{h}_k \mathbf{w}_k x_k + \sum_{j \in\mathcal{K} \backslash k} \mathbf{h}_k \mathbf{w}_j x_j + \sum_{l \in\mathcal{L}} \mathbf{g}_k \cdot \prod_{\substack{\ell=L, \ell > l,\\ \ell\leftarrow \ell-1}} (\boldsymbol{\Theta}_{\ell} \mathbf{H}_{\ell-1}) \cdot \boldsymbol{\Theta}_{l} \mathbf{n}_{l}  + n_k$, where $\mathbf{w}_k \in \mathbb{C}^{N\times 1}$ is the beamforming vector for user $k$, $x_k \sim \mathcal{CN}(0,1)$ is the transmitted symbol, $\mathbf{n}_{l} \sim \mathcal{CN}(\mathbf{0},\sigma_{\text{S}}^2\mathbf{I})$ indicates the noise of the $l$-th ASIM layer, and $n_k \sim \mathcal{CN}(0,\sigma^2)$ is user noise. The noise power of ASIM and user is denoted as $\sigma_{\text{S}}^2$ and $\sigma^2$, respectively. $\mathbf{I}$ is the identity matrix. The signal-to-interference-plus-noise ratio (SINR) of user $k$ can be given by
\begin{align}
	\gamma_k = \frac{|\mathbf{h}_k \mathbf{w}_k|^2}{\sum_{j \in\mathcal{K} \backslash k} |\mathbf{h}_k \mathbf{w}_j|^2 + I_{\text{S},k} + \sigma^2},
\end{align}
where $I_{\text{S},k}$ indicates the amplified noise of ASIM given by $I_{\text{S},k} = \sum_{l \in\mathcal{L}} \sigma_{\text{S}}^2 \lVert \mathbf{g}_k \cdot \prod_{\substack{\ell=L, \ell > l,\\ \ell\leftarrow \ell-1}} (\boldsymbol{\Theta}_{\ell} \mathbf{H}_{\ell-1}) \cdot \boldsymbol{\Theta}_{l} \rVert^2$. It is worth noting that each active ASIM layer not only amplifies the desired incident signal but also its inherent circuit noise. Consequently, the amplified noise generated at the $l$-th layer propagates through all subsequent ASIM layers before reaching the user. The achievable rate is defined as $R_k = \log_2(1 + \gamma_k)$. The total rate can be attained as $R_{\text{tot}} = \sum_{k\in\mathcal{K}} R_k$.

The total power consumption includes two parts: The BS consumes $P_{\text{B}} = 1/\eta_{\text{PA}} \sum_{k=1}^K \|\mathbf{w}_k\|^2 + P_{\text{fix}}$, where $\eta_{\text{PA}}$ indicates power amplifier efficiency, and $P_{\text{fix}}$ denotes the fixed power from operating circuits and baseband processing. Based on \cite{ris_THz}, the power consumption of ASIM can be derived by
\begin{align} \label{power_star}
	P_{\text{S}} = L \cdot M \cdot N_{\rm PIN} \cdot P_{\rm PIN} + L \cdot P_{\rm CIR} + \varsigma \cdot \sum_{l\in\mathcal{L}} P^{\text{O}}_{l},
\end{align}
where $N_{\rm PIN} = \lceil \log_2 Q_{\beta} + \log_{2} Q_{\theta} \rceil$ is the required number of PIN diodes for operating circuits controlling amplitude and phase-shifts. The notation $P_{\rm PIN}$ indicates the operating power consumption of each meta-element, whilst $P_{\rm CIR}$ indicates the static circuit power consumption of each ASIM. The quantization levels of amplitudes and phase-shifts are denoted as $Q_{\beta}$ and $Q_{\theta}$, respectively. Notation $\varsigma$ is the inverse of the power amplifier efficiency. The output power of the $l$-th ASIM is $P^{\text{O}}_{l}  =  \sum_{k\in \mathcal{K}} \lVert \prod_{\substack{\ell=l, \ell \geq 1,\\ \ell\leftarrow \ell-1}} \left( \boldsymbol{\Theta}_{\ell} \mathbf{H}_{\ell-1} \right) \mathbf{w}_{k}  \rVert^2 +
	\sigma_{\text{S}}^2 \cdot \sum_{\ell=1}^{l} 
 	\lVert \prod_{\substack{q=l-\ell+1, \\ q \leftarrow q-1, q\geq 2}} \left( \boldsymbol{\Theta}_{q} \mathbf{H}_{q-1} \right) \boldsymbol{\Theta}_{\ell} \rVert^2_{F}$, where $\left\lVert \cdot \right\rVert_{F}$ is the Frobenius norm. The output power consists of two components, i.e., amplified transmit signal power and accumulated amplified circuit noise power. The former represents the power required to amplify the incident signal, whereas the latter accounts for the unavoidable amplification of circuit noise introduced by the amplifiers. The system EE can be defined as $EE = \frac{R_{\text{tot}}}{P_{\text{B}}+ P_{\text{S}}}$. We aim for maximizing the total downlink rate by optimizing the BS beamforming $\mathbf{w}_k$ and configurations $\boldsymbol{\Theta}_{l}$ of ASIM, formulated as
\begingroup
\allowdisplaybreaks
\begin{subequations} \label{total_problem}
\begin{align}
& \max_{\substack{\mathbf{w}_k, \boldsymbol{\Theta}_l}} \quad EE \\
& \quad \text{ s.t.}  \quad
  	R_k \geq R_{\rm th}, \quad \forall k\in\mathcal{K}, \label{con1}\\
	& \qquad\quad \sum_{k\in\mathcal{K}} \|\mathbf{w}_k\|^2 \le P_{\max}, \label{con2} \\
	& \qquad\quad |\varphi_{l,m}|^2 \leq \beta_{\max}, \quad \forall l\in\mathcal{L}, \forall m \in \mathcal{M}. \label{con3}
\end{align}
\end{subequations}
\endgroup
Constraint \eqref{con1} guarantees the minimum rate requirement of each user $k$ by $R_{\rm th}$. In \eqref{con2}, maximum power is constrained by $P_{\max}$. \eqref{con3} stands for ASIM amplification constrained by $\beta_{\max}$. The problem is non-linear and non-convex, which is challenging. Therefore, we propose an AO technique to iteratively transform and solve the subproblems.

\section{Proposed AO-based Solution for ASIM}

\subsection{Problem Transformation}

Due to the fractional form of EE, we first adopt Dinkelbach method \cite{shen_dstar} by introducing the auxiliary variable $\eta$ to transform the problem, formulated as
\begingroup
\allowdisplaybreaks
\begin{subequations} \label{total_problem2}
\begin{align}
	& \max_{\substack{\mathbf{w}_k, \boldsymbol{\Theta}_l}, \eta} \quad R_{\text{tot}} - \eta( P_{\text{B}}+P_{\text{S}})  \label{obj_dl}\\
& \quad \text{s.t.}  \quad \eqref{con2}, \eqref{con3}, \quad \gamma_k \geq \gamma_{\rm th}, \quad \forall k\in\mathcal{K}. \label{con2_2}
\end{align}
\end{subequations}
\endgroup
In \eqref{obj_dl}, the optimum auxiliary variable $\eta^{\star} = EE$ is obtained by replacing the solutions at previous iteration, with the pertinent proof found in \cite{shen_dstar}. Note that $\gamma_{\rm th} = 2^{R_{\rm th}}-1$ in \eqref{con2_2} is alternated from \eqref{con1}. We further approximate $R_{\text{tot}}$ in \eqref{obj_dl} using Lagrangian dual transformation with the auxiliary parameter $\boldsymbol{\gamma}=\{\bar{\gamma}_k | \forall k\in\mathcal{K} \}$ as
\begin{align} \label{LD}
	f_{\text{LD}}(\mathbf{w},\boldsymbol{\Theta}) = \sum_{k\in \mathcal{K}} \log(1+\bar{\gamma}_k) -
\sum_{k\in \mathcal{K}} \bar{\gamma}_k \!+\!
\sum_{k\in \mathcal{K}} \frac{(1+\bar{\gamma}_k) A_k }{ B_k },
\end{align}
where $A_k = | \mathbf{h}_k \mathbf{w}_k|^2$ and $B_k = \sum_{j \in \mathcal{K}} |\mathbf{h}_k \mathbf{w}_j|^2 + I_{\text{S},k} + \sigma^2$. The optimum is $\bar{\gamma}_k^{\star} = \frac{| \mathbf{h}_k \mathbf{w}_k|^2}{\sum_{j \in \mathcal{K} \backslash k} |\mathbf{h}_k \mathbf{w}_j|^2 + I_{\text{S},k} + \sigma^2}$, where the proof can be found in \cite{shen_dstar}. Note that this term can be obtained from the solution at previous iteration, preserving equivalence with the original SINR expression. To this end, the last term in \eqref{LD} remains unsolved owing to the form of sum-of-ratios. By employing quadratic transform \cite{QT}, we can acquire
\begingroup
\allowdisplaybreaks
\begin{align} \label{y_lagrang}
	& f_{\text{QT}}(\mathbf{w},  \boldsymbol{\Theta}, \boldsymbol{\gamma}, \mathbf{y}) = \sum_{k\in\mathcal{K}} 2 \bar{y}_k \sqrt{ (1+ \bar{\gamma}_k) | \mathbf{h}_k \mathbf{w}_k|^2} \notag\\
	& \qquad\qquad -\sum_{k\in\mathcal{K}} \bar{y}_k^2 ( \sum_{j \in \mathcal{K}} |\mathbf{h}_k \mathbf{w}_j |^2 + I_{\text{S},k} + \sigma^2 ) + c_{\gamma},
\end{align}
\endgroup
where $\mathbf{y}=\{\bar{y}_k| \forall k\in\mathcal{K}\}$ and $c_{\gamma} = \sum_{k\in \mathcal{K}} \log(1+\bar{\gamma}_k) - 
\sum_{k\in \mathcal{K}} \bar{\gamma}_k$ is the constant term w.r.t. $\bar{\gamma}_k$, where the closed-form of $\bar{y}_k$ is derived as $\bar{y}_k^{\star} = \frac{\sqrt{ (1+\bar{\gamma}_k) | \mathbf{h}_k \mathbf{w}_k|^2 }}{\sum_{j \in \mathcal{K}} |\mathbf{h}_k \mathbf{w}_j|^2+ I_{\text{S},k}+ \sigma^2}$ by following \cite{QT}, preserves an asymptotic form to SINR. Note that this term is obtained from the solved solution at previous iteration. The transformed problem now becomes related to $\mathbf{w}_k$, $\boldsymbol{\Theta}_l$, and $\eta$. We adopt an AO scheme to solve the respective subproblems below.

\subsection{Solution to BS Beamforming}

Here, we fix $\boldsymbol{\Theta}_l$ and solve $\mathbf{w}_k$ and $\eta$. The subproblem is then reformulated as
\begin{align} \label{problem_w1}
	& \max_{\substack{\mathbf{w}_k}, \eta} \quad f_{w}(\mathbf{w}) - \eta ( P_{\text{B}}+P_{\text{S}}) \quad \text{ s.t. } \quad \eqref{con2}, \eqref{con2_2},
\end{align}
where $f_{w}(\mathbf{w})$ indicates the function $f_{\text{QT}}(\mathbf{w},  \boldsymbol{\Theta}, \boldsymbol{\gamma}, \mathbf{y})$ in \eqref{y_lagrang} w.r.t. $\mathbf{w}_k$ under the fixed $\boldsymbol{\Theta}_l$ and the fixed optimum solutions of $\bar{\gamma}_k^{\star}$ and $\bar{y}_k^{\star}$. The non-convex term in the objective lies on $|\mathbf{h}_k \mathbf{w}_k|$, which is approximated by utilizing SCA associated with the first-order Taylor expansion as
\begingroup
\allowdisplaybreaks
\begin{align} \label{g1}
	|\mathbf{h}_k \mathbf{w}_k^{(t)}| + \mathfrak{R}\{ ( \frac{\mathbf{h}_{k}^{\mathcal{H}} \mathbf{h}_{k} \mathbf{w}_{k}^{(t)} }{|\mathbf{h}_k \mathbf{w}_k^{(t)}|} )^{\mathcal{H}} (\mathbf{w}_k - \mathbf{w}_k^{(t)}) \} \triangleq \tilde{g}_1(\mathbf{w}_k),
\end{align}
\endgroup
where $t$ indicates the previous iteration index. Note that solving each convex SCA subproblem as first-order lower bounds produces a non-decreasing surrogate objective while preserving feasibility. Accordingly, the objective becomes 
\begingroup
\allowdisplaybreaks
\begin{align}
	\tilde{f}_{w}(\mathbf{w}) \!=\! \sum_{k\in\mathcal{K}} c_y \tilde{g}_1(\mathbf{w}_k) 
	\!-\! \sum_{k\in\mathcal{K}} \bar{y}_k^2 ( \sum_{j \in \mathcal{K}} |\mathbf{h}_k \mathbf{w}_j |^2 \!+\! I_{\text{S},k} ),
\end{align}
\endgroup
where $c_y = 2 \bar{y}_k \sqrt{ (1+ \bar{\gamma}_k) }$. Similarly, in \eqref{con2_2}, we adopt SCA with the first-order Taylor expansion to approximate the non-convex term $|\mathbf{h}_k \mathbf{w}_k|^2 \approx |\mathbf{h}_k \mathbf{w}_k^{(t)}|^2 +2\mathfrak{R}\{ (\mathbf{h}_k \mathbf{w}_k^{(t)})^* \mathbf{h}_k (\mathbf{w}_k \!-\! \mathbf{w}_k^{(t)})\} \triangleq \tilde{g}_2(\mathbf{w}_k)$. Therefore, $\eqref{con2_2}$ becomes
\begin{align} \label{sca_gamma}
	\tilde{g}_2(\mathbf{w}_k) \geq \gamma_{\rm th} \cdot ( \sum_{j \in\mathcal{K} \backslash k} |\mathbf{h}_k \mathbf{w}_j|^2 + I_{\text{S},k} + \sigma^2 ).
\end{align}
The subproblem is then reformulated as
\begin{align} \label{problem_w2}
	& \max_{\substack{\mathbf{w}_k, \eta}} \quad \tilde{f}_{w}(\mathbf{w}) - \eta ( P_{\text{B}}+P_{\text{S}}) \quad \text{ s.t. } \quad \eqref{con2}, \eqref{sca_gamma},
\end{align}
which is convex and solved by any convex optimization tools.

\subsection{Solution to ASIM Configuration}
	 
Now, we solve the subproblem w.r.t. $\boldsymbol{\Theta}_l$ by fixing $\mathbf{w}_k$ and $\eta$, reformulated as
\begin{align} \label{problem_t1}
	& \max_{\substack{\boldsymbol{\Theta}_l}} \quad f_{\theta}(\boldsymbol{\Theta}) - \eta ( P_{\text{B}}+P_{\text{S}}) \quad \text{ s.t. } \quad \eqref{con3}, \eqref{con2_2},
\end{align}
where $f_{\theta}(\boldsymbol{\Theta})$ indicates the function $f_{\text{QT}}(\mathbf{w},  \boldsymbol{\Theta}, \boldsymbol{\gamma}, \mathbf{y})$ in \eqref{y_lagrang} w.r.t. $f_{\theta}(\boldsymbol{\Theta})$ under the fixed $\mathbf{w}_k$, $\bar{\gamma}_k^{\star}$ and $\bar{y}_k^{\star}$. Since the configurations are highly coupled through the cascaded near-field channel in \eqref{ch_general}, iterative layer-wise optimization is prone to converging to local solutions. Furthermore, gradient-based methods, such as projected gradient and manifold optimization, require analytical gradients or repeated gradient evaluations, whose computational cost can increase with the coupled multi-layer variables. Therefore, Bayesian optimization \cite{BO} is adopted to jointly optimize the ASIM configuration in a gradient-free manner. Let us define the ASIM optimization vector as $\mathbf{x} = [\boldsymbol{\beta}_{1}^{\mathcal{T}},..., \boldsymbol{\beta}_{L}^{\mathcal{T}}, \boldsymbol{\theta}_{1}^{\mathcal{T}},..., \boldsymbol{\theta}_{L}^{\mathcal{T}}
]$, where $\mathcal{T}$ is transpose operation. Note that $\boldsymbol{\beta}_{l}=[\beta_{l,1},...,\beta_{l,M}]^{\mathcal{T}}$ and $\boldsymbol{\theta}_{l}=[\theta_{l,1},...,\theta_{l,M}]^{\mathcal{T}}$. Then the objective function for Bayesian optimization is defined as
\begin{align} \label{F_bo}
 F(\mathbf{x}) = f_{\theta}(\boldsymbol{\Theta}) - \eta ( P_{\text{B}}+P_{\text{S}}) - \sum_{k\in\mathcal{K}} \lambda_k \cdot [\gamma_{\rm th}- \gamma_k]^+,
\end{align}
where $[z]^+=\max(z,0)$ and $\lambda_k$ is the penalty coefficient for the SINR constraint. This converts the constrained ASIM configuration problem into an unconstrained black-box maximization problem. Bayesian optimization then constructs a Gaussian process (GP) surrogate model as $F(\mathbf{x}) \sim \mathcal{GP}({\mu}(\mathbf{x}), k(\mathbf{x}, \mathbf{x}'))$, where ${\mu}(\mathbf{x})$ is the posterior mean and $ k(\mathbf{x}, \mathbf{x}')$ is the kernel function. At iteration $t$, given the observed dataset $\mathcal{D}_t=\{ (\mathbf{x}_i, z_i) \}_{i=1}^t$ where $z_i=F(\mathbf{x}_i)$. The GP posterior mean and variance for a new ASIM candidate are respectively derived by
\begingroup
\allowdisplaybreaks
\begin{align}
	\mu_t(\mathbf{x}) &= m(\mathbf{x}) +\mathbf{k}_t^{\mathcal{T}}(\mathbf{x}) \left( \mathbf{K}_t +\sigma_n^2 \mathbf{I} \right)^{-1} (\mathbf{z}_t - \mathbf{m}_t), \\
	\sigma_t^2(\mathbf{x}) &= k(\mathbf{x}, \mathbf{x}) - \mathbf{k}_t^{\mathcal{T}} \left( \mathbf{K}_t +\sigma_n^2 \mathbf{I} \right)^{-1} \mathbf{k}_{t}(\mathbf{x}), 
\end{align}
\endgroup
where $\mathbf{z}_t=[z_1, ..., z_t]^{\mathcal{T}}$, $\mathbf{k}_t(\mathbf{x}) = [k(\mathbf{x}, \mathbf{x}_1),..., k(\mathbf{x}, \mathbf{x}_t)]^{\mathcal{T}}$, and $\mathbf{K}_t =[k(\mathbf{x}_i, \mathbf{x}_j)]_{i,j=1}^t$. Note that $m(\mathbf{x})= \mathbbm{E}[F(\mathbf{x})]$ and $\mathbf{m}_t = [m(\mathbf{x}_1),..., m(\mathbf{x}_t)]^{\mathcal{T}}$. To explicitly connect the kernel with $\beta_{l,m}$ and $\theta_{l,m}$, we employ $k(\mathbf{x}, \mathbf{x}') = \sigma_f^2 e^{-\frac{1}{2} d^2(\mathbf{x}, \mathbf{x}')}$, where $d^2(\mathbf{x}, \mathbf{x}') \!=\! \sum_{l\in\mathcal{L}} \sum_{m\in\mathcal{M}} \frac{(\beta_{l,m} \!-\! \beta_{l,m}')^2}{\ell^2_{\beta}} \!+\! \frac{|e^{j\theta_{l,m}} \!-\! e^{j\theta_{l,m}'}|^2}{\ell^2_{\theta}}$. Note that $\sigma_n^2$ indicates observation noise variance, $\sigma_f^2$ means signal variance of the GP kernel, and $\ell_{\beta},\ell_{\theta}$ represent the kernel length-scales for the ASIM amplification and phase-shift, respectively. The next ASIM configuration is selected based on the designed acquisition function as
\begin{align} \label{GPfit}
	\mathbf{x}_{t+1} = \arg\max_{\mathbf{x} \in \mathcal{X}} \ \mu_t(\mathbf{x})+\varpi_t \sigma_t (\mathbf{x}),
\end{align}
where $\mathcal{X} = \{\mathbf{x}|0\leq\beta_{l,m}\leq \beta_{\max}, 0\leq \theta_{l,m}< 2\pi \}$. Notation of $\mu_t(\mathbf{x})$ makes the predicted ASIM objective value, where $\sigma_t (\mathbf{x})$ indicates the posterior uncertainty for each candidate configuration. $\varpi_t$ controls the exploration-exploitation tradeoff. The acquisition function \eqref{GPfit} is numerically maximized over the feasible set $\mathcal{X}$. Owing to its inexpensive evaluation through GP posterior mean and variance, function of \eqref{GPfit} can be efficiently optimized by a multi-start local search. The selected candidate is subsequently evaluated using the original EE objective, and the resulting observation is incorporated into the GP model to update the posterior distribution for the next Bayesian optimization iteration. We further note that the final Bayesian solutions of configurations are uniformly quantized according to the available hardware resolutions before being applied to the active meta-atoms.

The detailed algorithm is elaborated in Algorithm \ref{alg}. Note that the actual runtime depends heavily on the implementation platform, e.g., the processors, optimization solvers, and software environments. Therefore, we provide a detailed computational complexity analysis of the proposed AO algorithm as follows. The total number of beamforming optimization variables is $KN_t$. Based on the convex subproblem in \eqref{problem_w2} solved by the interior-point method, its computational complexity is attained as $\mathcal{O}\left(I_{\rm SCA}(KN_t)^{3.5}\right)$, where $I_{\rm SCA}$ denotes the number of SCA iterations. For ASIM subproblem, the optimization is with the dimension of $D_{\rm ASIM}=2LM$. In Bayesian optimization given $N_{\rm BO}$ evaluated samples, the covariance matrix inversion requires $\mathcal{O}(N_{\rm BO}^3)$. Moreover, computing the mean-variance functions over $N_{\rm sol}$ candidates requires $\mathcal{O}(N_{\rm sol}N_{\rm BO}D_{\rm ASIM})$. Accordingly, complexity of ASIM subproblem is derived by
$ \mathcal{O} \left( I_{\rm BO} \left( N_{\rm BO}^3 + N_{\rm sol}N_{\rm BO}D_{\rm ASIM} \right) \right)$, where $I_{\rm BO}$ denotes the number of Bayesian iterations. Accordingly, the total complexity of the proposed solution is $\mathcal{O} \left( I_{\rm AO} \left[ I_{\rm SCA}(KN_t)^{3.5} + I_{\rm BO} \left( N_{\rm BO}^3 + N_{\rm sol}N_{\rm BO}D_{\rm ASIM} \right) \right] \right)$, where $I_{\rm AO}$ is the number of outer AO iterations.

%
%

\begin{algorithm}[!tb]
\small
\caption{Proposed AO-based Solution for ASIM}
\SetAlgoLined
\DontPrintSemicolon
\label{alg}
\begin{algorithmic}[1]
\STATE {\bf Initialization:} Set $t=0$, convergence tolerance $\epsilon$, maximum iteration $T_{\max}$, $\mathbf{w}_k^{(0)}$, and $\boldsymbol{\Theta}_l^{(0)}$; $\eta^{(0)}=R_{\rm tot}^{(0)}/ (P_B^{(0)}+P_S^{(0)})$
\REPEAT
    \STATE $t\leftarrow t+1$, compute channel \eqref{ch_general}, and update auxiliaries\\ $\bar{\gamma}_k^{(t)}$, $\bar{y}_k^{(t)}$
    \STATE {\bf (BS Beamforming Optimization)}
    \STATE Initialize $t_1=0$
    \REPEAT 
    	\STATE $t_1\leftarrow t_1+1$ and construct $\tilde{g}_1(\mathbf{w}_k)$, $\tilde{g}_2(\mathbf{w}_k)$ in \eqref{g1}, \eqref{sca_gamma}
    	\STATE Solve the convex beamforming subproblem in \eqref{problem_w2}
	\UNTIL {Convergence}
	\STATE {\bf (ASIM Configuration Optimization)}
	\STATE Initialize $t_2=0$ and dataset $\mathcal{D}_0=\emptyset$
    \REPEAT
    	\STATE $t_2\leftarrow t_2+1$
    	\STATE  Define $\mathbf{x}$, fit GP model using $\mathcal{D}_{t-1}$, and compute\\ $\mu_t(\mathbf{x})$, $\sigma_t(\mathbf{x})$
    	\STATE Select next ASIM configuration by \eqref{GPfit} and reconstruct $ \phi_{l,m}^{(t)}$
    	\STATE  Evaluate $F(\mathbf{x}_{t}^{\star})$ in \eqref{F_bo} and  update dataset
    $
    \mathcal{D}_{t}
    \!=\!
    \mathcal{D}_{t\!-\!1}
    \cup
    \{(\mathbf{x}_{t}^{\star},F(\mathbf{x}_{t}^{\star}))\}
    $
    \UNTIL {Convergence}
    \STATE Update
    $
    \eta^{(t)}
    =
    R_{\text{tot}}^{(t)} / (P_{\text{B}}^{(t)}+P_{\text{S}}^{(t)})
    $
\UNTIL{$|\eta^{(t)}-\eta^{(t-1)}|\leq \epsilon$ or $t\geq T_{\max}$}
\STATE {\bf Output:} Optimized beamforming $\mathbf{w}_k^\star$, ASIM configurations $\boldsymbol{\Theta}_l^\star$, and EE $\eta^\star$
\end{algorithmic}
\end{algorithm}

\begin{table}[t]
\centering
\small
\caption{QoS Violation Probability}
\label{qos_vio}
\begin{tabular}{|c|c|c|c|c|}
\hline
QoS $R_{\rm th}$ (bps/Hz)
& $0.5$ & $1$ & $1.5$ & $2$ \\
\hline
Violation Probability
& $0$ & $0.08$ & $0.16$ & $0.27$ \\
\hline
\end{tabular}
\end{table}

\begin{figure*}[!t]
	\centering
	\subfigure[]{\includegraphics[width=3.5 in]{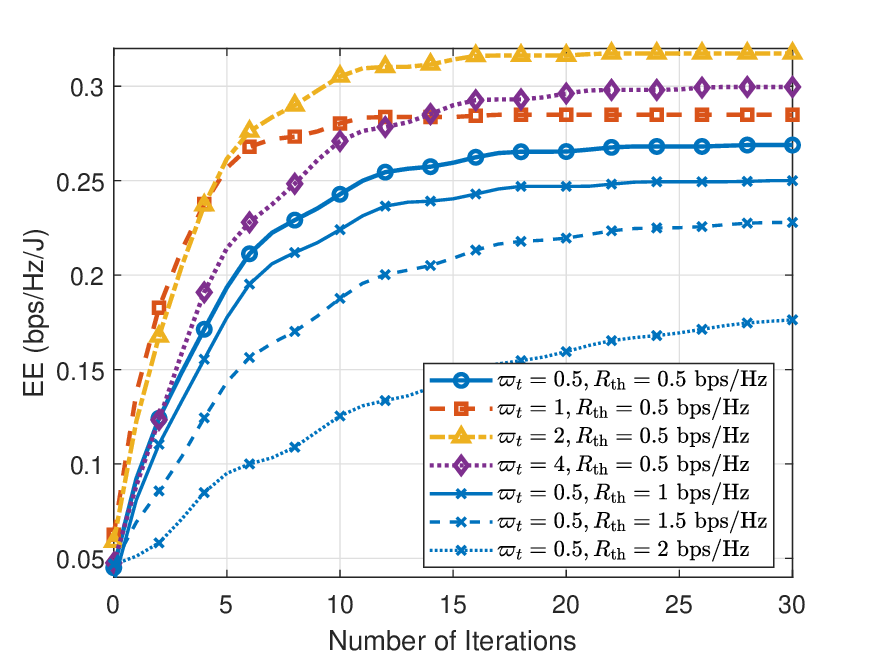} \label{converge}}
	\subfigure[]{\includegraphics[width=3 in]{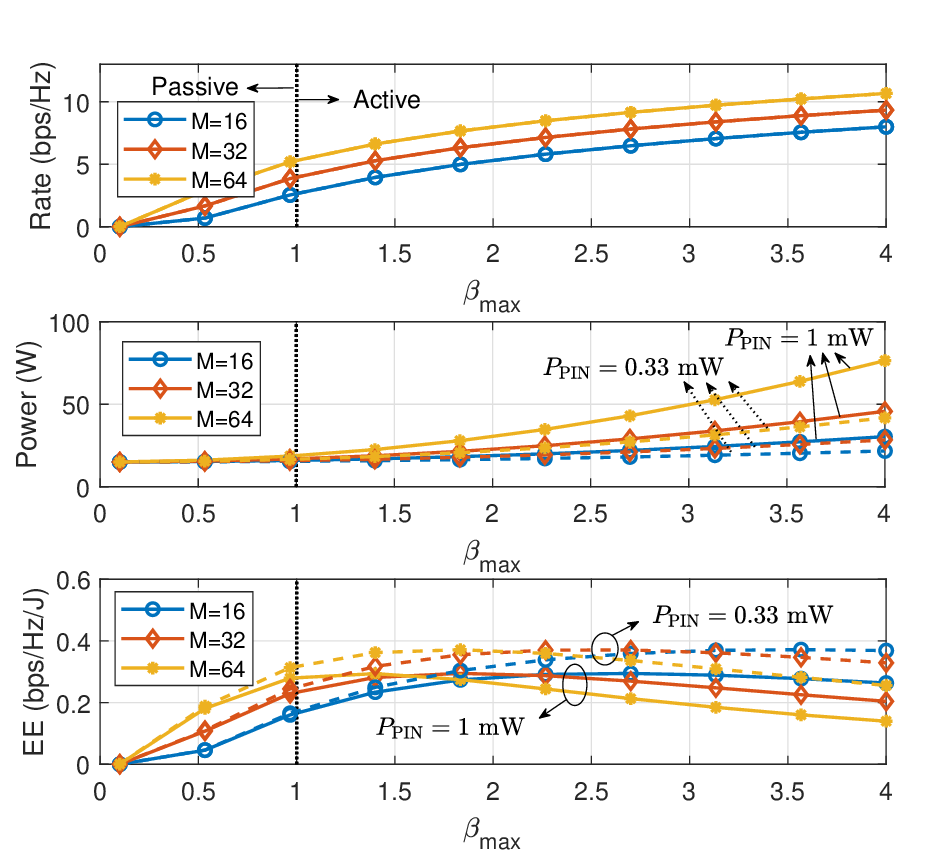} \label{r1}}
	\subfigure[]{\includegraphics[width=3.5 in]{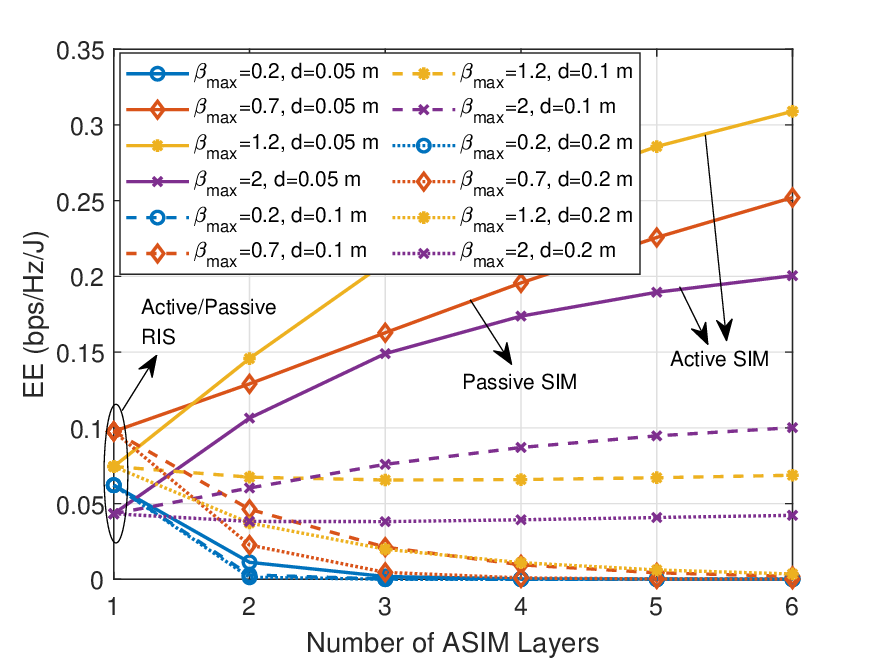} \label{r2}}
	\subfigure[]{\includegraphics[width=3.5 in]{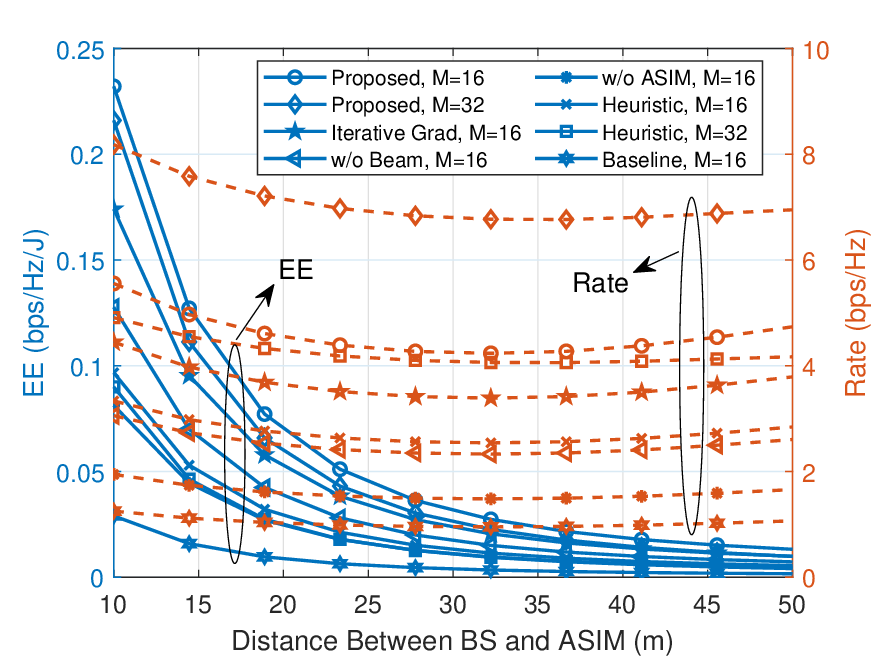} \label{r3}}
	\caption{
	(a) Convergence.
	(b) Rate/power/EE of ASIM with different $\beta_{\rm max}$, $M$ and $P_{\rm PIN}$. 
	(c) EE with different $\beta_{\rm max}$, $d$ and $L$. 
	(d) Benchmark comparison.
	}
\end{figure*}

\section{Simulation Results}


Simulation results are provided to validate the proposed ASIM-assisted downlink network. The BS is located at the origin and serves $K=4$ single-antenna users which are independently and uniformly distributed within a circular region of radius 5 m centered at $(50, 0, 1.5)$ m. The ASIM is deployed at $(20,0,5)$ m and consists of $L=3$ layers with each layer having $M=32$ elements. All results are averaged over $100$ independent Monte Carlo channel realizations. The carrier frequency is $3.5$ GHz, the pathloss exponent is $\alpha_k=2.2$ \cite{3gpp}, and all wireless links follow Rician fading with $\kappa=\kappa_k=3$. The BS employs $N_t=8$ antennas. The inter-layer spacing is $d=0.05$ m, while the element area and inter-element spacings are set to $A_m=\lambda^2/4$ and $d_{\text{S}}=d_{\text{B}}=\lambda/2$, respectively. The noise power is $\sigma^2=\sigma_{\text{S}}^2=-80$ dBm, the maximum transmit power is $P_{\max}=30$ dBm, and the minimum rate requirement is $R_{\rm th}=0.5$ bps/Hz. Note that $\eta_{\text{PA}}=0.4$ and $P_{\text{fix}}=5$ Watt. The ASIM amplification limit is $\beta_{\max}=2$. The power consumption parameters are $P_{\rm PIN}=0.33$ mW, $P_{\rm CIR}=10$ W, and $\varsigma=1.2$ \cite{ris_THz}, while quantization levels are $Q_{\beta}=Q_{\theta}=16$, which provide performance close to the continuous-control upper bound while maintaining practical hardware complexity. For the proposed algorithm, $\epsilon=10^{-3}$, $T_{\max}=30$, and $\lambda_k=10$. In Bayesian optimization, $N_{\rm BO}=20$, $I_{\rm BO}=50$, $\sigma_n^2=10^{-6}$, $\sigma_f^2=1$, $\ell_{\beta}=1$, and $\ell_{\theta}=\pi/2$. 

Fig. \ref{converge} depicts the convergence of the proposed scheme under different exploration rate $\varpi_t$. A small exploration coefficient $\varpi_t=0.5$ leads to premature convergence and lower EE due to insufficient exploration, whereas $\varpi_t=4$ slows convergence because of excessive exploration. The case with $\varpi_t=2$ achieves the highest converged EE while maintaining a fast convergence rate, providing the best exploration-exploitation tradeoff, which is adopted in the following simulations. We further observe that a more stringent QoS requirement $R_{\rm th}$ reduces the achievable EE, since additional transmit and amplification resources are required to satisfy the minimum-rate constraints. Moreover, the QoS violation probability increases with higher $R_{\rm th}$, as elaborated in Table \ref{qos_vio}.

Fig. \ref{r1} illustrates the impacts of the maximum amplification coefficient $\beta_{\max}$. The achievable rate monotonically increases with $\beta_{\max}$ for all cases. In the passive regime ($\beta_{\max}\leq 1$), the performance gain mainly stems from the enlarged aperture provided by more ASIM elements, whereas active amplification ($\beta_{\max}>1$) further enhances the received signal strength and rate. Meanwhile, the power consumption increases almost linearly with $\beta_{\max}$ due to the additional amplification power, and the growth becomes more pronounced for larger ASIMs. Moreover, $P_{\rm PIN}=1$ mW incurs higher power consumption than $P_{\rm PIN}=0.33$ mW, with the gap enlarging as $\beta_{\max}$ increases. Consequently, the EE exhibits a concave trend with respect to $\beta_{\max}$. Specifically, EE first increases as the rate gain outweighs the power expenditure, but eventually decreases when the marginal rate improvement becomes insufficient to compensate for the additional power consumption. The results suggest that, under the considered practical ASIM hardware settings, a moderate amplification capability provides the most favorable EE tradeoff between achievable rate and power consumption. Furthermore, the lower-power hardware setting of $P_{\rm PIN}=0.33$ mW consistently achieves higher EE, highlighting the importance of energy-efficient circuit design for active metasurfaces.

Fig. \ref{r2} illustrates the impact of the number of ASIM layers under different $\beta_{\max}$ and $d$. Note that the case of $L=1$ and $\beta_{\max}\leq 1$ corresponds to the single-layer passive RIS \cite{shen_amy}, that of $L=1$ and $\beta_{\max}>1$ corresponds to active RIS \cite{a_ris}, while that of $L>1$ and $\beta_{\max}\leq 1$ represents the passive SIM \cite{sim1, eesim}. The remaining cases $L>1$ and $\beta_{\max} > 1$ correspond to the proposed ASIM. For the case of $d=0.05$ m, the EE generally increases with the number of layers when $\beta_{\max}\geq 0.7$. This is because additional ASIM layers provide more degrees of freedom for electromagnetic wave manipulation and beam focusing, enhancing the rate. Particularly, $\beta_{\max}=1.2$ achieves the highest EE among all settings, indicating that a moderate amplification effectively balances rate and power consumption. However, when $\beta_{\max}=2$, the EE growth becomes less significant owing to excessive power consumed by active circuits. For larger inter-layer spacings of $d\in\{0.1,0.2\}$ m, EE deteriorates with more layers. This can be attributed to the weakened near-field coupling among adjacent layers, reducing beamforming gains obtained from stacked layers. Furthermore, the superiority of smaller inter-layer spacing becomes more evident when the amplification capability increases. On the other hand, for the passive case $\beta_{\max}=0.2$, increasing layers provides limited EE gains regardless of inter-layer spacing due to the absence of sufficient amplification capability.

Fig. \ref{r3} depicts the impacts of the BS-ASIM distance under different benchmarks. It can be observed that EE decreases as the distance increases. This is expected since a larger BS-ASIM distance introduces more severe propagation loss in the BS-ASIM link, reducing the effective channel gain and limiting the benefits of ASIM. Specifically, the rate initially decreases with the BS-ASIM distance due to the increasing propagation loss. However, a slight rate recovery is observed at larger distances because the shortened ASIM-user distance improves the second-hop channel quality. Consequently, the enhanced ASIM-user links partially compensate for the degradation of the BS-ASIM channel through adaptive beamforming and ASIM configuration. Moreover, the proposed scheme achieves the highest EE and rate across all distances. This superiority stems from the joint optimization of BS beamforming and ASIM configurations, fully exploiting the available spatial and amplification gains. In contrast, "w/o Beam" indicating non-optimized beamforming suffers from a noticeable performance degradation. Similarly, "w/o ASIM" with non-optimized configurations exhibits the lowest performance as the propagation cannot be intelligently reconfigured. Comparing different benchmarks, the proposed AO-based Bayesian optimization approach outperforms both the iterative gradient-based optimization \cite{sim_iter}, metaheuristic \cite{ga}, and random baseline methods, demonstrating the effectiveness of the Bayesian optimization framework in exploring highly-coupled ASIM configuration space and avoiding local solutions.

\section{Conclusions}
In this paper, we have investigated an energy-efficient ASIM-assisted downlink transmission framework, where a multi-antenna BS jointly optimizes transmit beamforming and multi-layer ASIM configurations. The resulting EE maximization problem is formulated under transmit power, amplification, and QoS constraints. To address the highly-coupled and non-convex problem, we have conceived an AO solution, where the BS beamforming subproblem is solved via Dinkelbach and SCA, whilst ASIM configuration subproblem is handled through the Bayesian optimization algorithm. Numerical results have demonstrated that the proposed solution significantly improves EE compared to conventional passive SIM and benchmarks. Furthermore, the results also strike important tradeoffs among the amplification capability, number of ASIM layers, inter-layer spacing, deployment location, and hardware power consumption. Particularly, properly configuring ASIM can substantially outperform passive counterparts by effectively balancing communication performance and power expenditure.

\bibliographystyle{IEEEtran}
\bibliography{IEEEabrv}
\end{document}